\documentclass[english,A4]{IEEEtran}
\usepackage[T1]{fontenc}
\usepackage[latin9]{inputenc}
\usepackage{amsthm}
\usepackage{amstext}
\usepackage{amssymb}
\usepackage{graphicx}
\usepackage{enumerate}
\usepackage{mathtools}
\usepackage{amsmath}
\usepackage{graphicx}
\usepackage{cuted}
\usepackage{flushend}
\usepackage{eucal}
\usepackage{pdflscape}
\usepackage{breqn}
\usepackage{multirow}
\newtheorem{theorem}{Theorem}
\newtheorem{lemma}{Lemma}

\makeatletter
\theoremstyle{definition}

\theoremstyle{definition}

\theoremstyle{definition}
\newtheorem{defn}{\protect Definition}
\theoremstyle{definition}
\newtheorem{rmrk}{\protect Remark}

\makeatother

\usepackage{babel}

\begin{document}

\title{Robust Gaussian Joint Source-Channel Coding with a Staircase Distortion-Noise Profile}

\author{Mohammadamin Baniasadi\\
University of California, Riverside, CA. Email: mohammadamin.baniasadi@email.ucr.edu,
}
\maketitle
\thispagestyle{empty}

\begin{abstract}
Minimum energy required to achieve a distortion-noise profile, i.e., a function indicating the maximum allowed distortion value for each channel noise level.
In this paper, the minimum energy required to achieve a distortion noise profile is studied for Gaussian sources which are transmitted robustly over Gaussian channels. 
We provide upper bound for the minimum energy behavior of the staircase profile using our proposed coding scheme.
Conversely, utilizing a family of lower bounds originally derived for broadcast channels with power constraints, the minimum required energy is lower bounded for staircase profile.
\end{abstract}

\section{Introduction}

Lossy transmission of source signals over noisy channels, which is in general a joint source-channel coding (JSSC) problem is requied in most of emerging wireless applications, such as Internet of things (IoT) and multimedia streaming. Shannon proved the separation theorem which indicates that in point-to-point scenarios, it is optimal to separate source and channel coding problems. However, in many problems, the optimality of separation breaks down, since JSCC can exploit source correlation to generate correlated channel inputs despite the distributed nature of the encoders, potentially improving the overall performance \cite{c1}-\cite{c4}.

We study lossy transmission of a Gaussian source over an additive white Gaussian noise (AWGN) channel, where the channel input constraint is on {\em energy per source symbol}. 
This regime has become more popular recently, such as \cite{c5,c6,c7, c8} to name as some references.
One important aspect is the simplifications to both achievable schemes and converses as the bandwidth expansion factor approaches infinity~\cite{c7}.

For a fixed channel noise variance $N$, it is well-known (for example, see~\cite{c5}) that the minimum distortion that can be achieved with  energy $E$ is given by
\begin{equation}
\label{eqtn:DvsE}
D = \exp\left(-\frac{E}{N}\right) \; .
\end{equation}

In this paper, we instead consider that $N$ can take any value in the interval $\left(0,\infty\right)$. This setting is robust which means $N$ is unknown at the transmitter (but known at the receiver as usual).  
The system is to be designed to fulfill with a distortion-noise profile ${\cal D}(N)$ so that it achieves 
\[
D \leq {\cal D}(N) 
\]
for all $0<N<\infty$, while minimizing its energy use. 
We consider this wide spectrum of noise variances to account for the scenarios that we may know {\em absolutely nothing} about the noise level. 
For instance, even though the channel may be originally of very high quality ($N\approx 0$), it could be suffering occasional interferences of a wide spectrum of noise levels (including $N\gg 0$). 

In \cite{c9}, it is shown that for the inversely linear profile, uncoded transmission is optimal. Furthermore, it is represented that exponential profiles are not achievable with finite energy. The square-law profile is also studied which is somehow combination of linear and exponential profiles and lower and upper bounds have been derived for the minimum achievable energy of the square-law profile.

In \cite{AminArxiv}, we derived improved lower and upper bounds for the minimum energy, and showed that the gap between our lower and upper bounds is significantly reduced compared to \cite{c9}. Improving lower and upper bounds and making them as tight as possible helps us to design better systems in practical scenarios
by comparing the amount of energy with these improved theoretical bounds.

Both \cite{c9} and \cite{AminArxiv} are in the context of infinite bandwidth. In \cite{AminArxivISIT}, we addressed the other extreme, where the bandwidth is severely limited. This near-zero bandwidth condition might arise in cases where too many devices (e.g., in Internet-of-Things networks) share the same communication medium through multiplexing (e.g., TDMA, FDMA, etc.)

One of the similar universal coding scenarios in the literature is given in \cite{c10}, where a maximum regret approach for compound channels is proposed. The objective in their scenario is to minimize the maximum ratio of the capacity to the achieved rate at any noise level.
Other related works include \cite{c11}, \cite{c12}, and \cite{c13}.

In this paper, we study staircase profile which is a practical profile and establish upper and lower bounds on the minimum energy for this profile. The rest of the paper is organized as follows. The next section is devoted to notation and preliminaries. In section~III, previous work on lower and upper bounds for the minimum energy is reviewed. In Section~IV, we present our main results, which are lower and upper bounds for the staircase profile. Finally, in Section~V we conclude our work and discuss future work.
\section{Notation and Preliminaries}
Suppose that $X^n$ is an i.i.d unit-variance Gaussian source which is transmitted over an AWGN channel $V^m=U^m+W^m$, where $U^m$ is the channel input, $W^m\sim \mathcal{N}(\mathbf{0},N\mathbf{I_m})$ is the noise, and $V^m$ is the observation at the receiver. We define bandwidth expansion factor $\kappa=\frac{m}{n}$ which can be arbitrarily large, while the energy per source symbol is limited by 
\begin{equation}
\label{eqtn:Energy}
\frac{1}{n} \mathbb{E}\left\{||U^m||^2\right\} \leq E \; .
\end{equation}
The achieved distortion per source symbol is measured as
\begin{equation}
\label{eqtn:Distortion}
D=\frac{1}{n}\mathbb{E}\left\{||X^n-\hat{X}^n||^2\right\}
\end{equation}
where $\hat{X}^n$ is the reconstruction at the receiver. 
\begin{defn}
\label{defn:Achievability}
A pair of distortion-noise profile ${\cal D}(N)$ and energy level $E$ is said to be {\em achievable} if for every $\epsilon>0$, there exists large enough $(m,n)$, an encoder
\[
f^{m,n}: \mathbb{R}^n\longrightarrow \mathbb{R}^m \; ,
\]
and decoders
\[
g^{m,n}_{N}: \mathbb{R}^m\longrightarrow \mathbb{R}^n
\]
for every $0<N<\infty$, such that
\[
\frac{1}{n} \mathbb{E}\left\{||f^{m,n}(X^n)||^2\right\} \leq E+\epsilon
\]
and
\[
\frac{1}{n}\mathbb{E}\left\{||X^n-g^{m,n}_{N}(f^{m,n}(X^n)+W^m_{N})||^2\right\} \leq {\cal D}(N)+\epsilon
\]
for all $N$, with $W^m_{N}$ being the i.i.d.\ channel noise with variance $N$. 
\end{defn}

For given ${\cal D}$, the main quantity of interest would be
\[
E_{\min}({\cal D}) = \inf \{E : ({\cal D},E) \mbox{ achievable}\} 
\]
with the understanding that $E_{\min}({\cal D}) =\infty$ if there is no finite $E$ for which $({\cal D},E)$ is achievable.

It will be much more convenient to use the notation $F = \frac{1}{D}$ and $Q = \frac{1}{N}$, $F$ and $Q$ standing for signal {\em fidelity} and channel {\em quality}\footnote{We cannot use the usual channel SNR as a quality measure since for any finite energy $E$, the expended power per channel symbol $\frac{E}{\kappa}$ approaches 0 as $\kappa\rightarrow\infty$.}, respectively. For any ${\cal D}(N)$, we define the corresponding {\em fidelity-quality profile} as
\[
{\cal F}(Q) = \frac{1}{{\cal D}(\frac{1}{Q})}
\]
and state that $(\mathcal{F},E)$ is achievable if and only if $(\mathcal{D},E)$ is achievable according to Definition~\ref{defn:Achievability}. $E_{\min}({\cal F})$ is similarly defined.

\section{Previous Work}
\subsection{A Family of Lower Bounds on $E_{min}(\mathcal{D})$}
In \cite{c9}, the authors used the connection between the problem and lossy transmission of Gaussian sources over Gaussian broadcast channels where the power per channel symbol is limited and the bandwidth expansion factor $\kappa$ is fixed. More specifically, they employed the converse result by Tian \textit{et al.} \cite{c14}, which is a generalization of the 2-receiver outer bound shown by Reznic \textit{et al.} \cite{c15} to $K$ receivers, and proved the following lemma.

\begin{lemma}
\label{Lemma1}
For any $K$, $\tau_1 \ge \tau_2 \ge ...\ge \tau_{K-1} \ge \tau_{K}=0$, and $N_1 \ge N_2 \ge ... \ge N_{K}\ge N_{K+1}=0$,
\begin{eqnarray} 
\label{eqtn:lemma1}
E_{min}(\mathcal{D})\!\!\!&\ge & \!\!\!N_1 \log \frac{1+\tau_1}{\mathcal{D}(N_1)+\tau_1}\nonumber\\
\!\!\!&& \!\!\!+\sum_{k=2}^{K} N_k \log \frac{(1+\tau_k)(\mathcal{D}(N_k)+\tau_{k-1})}{(1+\tau_{k-1})(\mathcal{D}(N_k)+\tau_{k})}.
\end{eqnarray}
\end{lemma}

\subsection{Square-Law Fidelity Quality Profiles}
In \cite{c9}, the authors focused on $\mathcal{F}(Q)=1+\alpha Q^2$ for some $\alpha >0$ and analyzed the lower and upper bounds for $E_{min}{(\mathcal{D})}$.
\subsubsection{Lower Bound for $E_{min}(\mathcal{D})$}
Invoking Lemma 1 by properly choosing $\tau_k$ and $N_k$ in (\ref{eqtn:lemma1}), the following theorem was obtained in \cite{c9}.

\begin{theorem}
For a fidelity-quality profile $\mathcal{F}(Q)=1+\alpha Q^2$, the minimum required energy is lower-bounded as
\begin{align}
E_{min}(\mathcal{D}) \ge c \ \sqrt[]{\alpha}\nonumber
\end{align}
with
\begin{align}
c=\sum_{k=1}^{\infty}\frac{1}{\sqrt[]{4^k\exp(k)-1}}\approx 0.4507. \nonumber
\end{align}
\end{theorem}
The Lemma 1 is general and works for any profile. However, it is not guaranteed that this lemma always gives us the best lower bound. For example, in \cite{AminArxiv} we could find better lower bound for a fidelity-quality profile $\mathcal{F}(Q) = 1+\alpha Q^2$.
\subsubsection{Upper Bound for $E_{min}(\mathcal{D})$}
Using a scheme first sending the source uncoded, and leveraging the received output as side information for the subsequent digital rounds sending indices of an infinite-layer quantizer, an upper bound for the minimum energy was presented in the following theorem in \cite{c9}.

\begin{theorem}
The minimum required energy for profile $\mathcal{F}(Q)=1+\alpha Q^2$ is upper-bounded as
\begin{align}
E_{min}(\mathcal{D})\leq d \ \sqrt[]{\alpha}\nonumber
\end{align}
with 
\begin{align}
d=2 \ \sqrt[]{\log3-Li_2(-2)}\approx 3.1846 \nonumber
\end{align}
where $Li_2(.)$ is the polylogarithm of order 2 defined as 
\begin{align}
Li_2(z)=-\int_{0}^{1} \frac{\log(1-zu)}{u}du.\nonumber
\end{align}
\end{theorem}
In \cite{AminArxiv}, we could improve this upper bound significantly. In \cite{AminArxiv}, instead of relying on only one uncoded transmission of the source $X^n$ as the generator of the side information at the receiver, we also send quantization errors uncoded after each layer of quantization. In other words, we have $K$ layers of uncoded transmission while in \cite{c9} the authors only had the
uncoded transmission in first layer.
\section{Analysis for Staircase Profile}
In this section, we focus on staircase profile. As an example, a staircase profile for $K=2$ is showed in Figure 1. Since it is difficult to look at this profile in its full generality, we will focus on two especial cases as follows.
\begin{figure}
  \centering
    \includegraphics[width=0.5\textwidth]{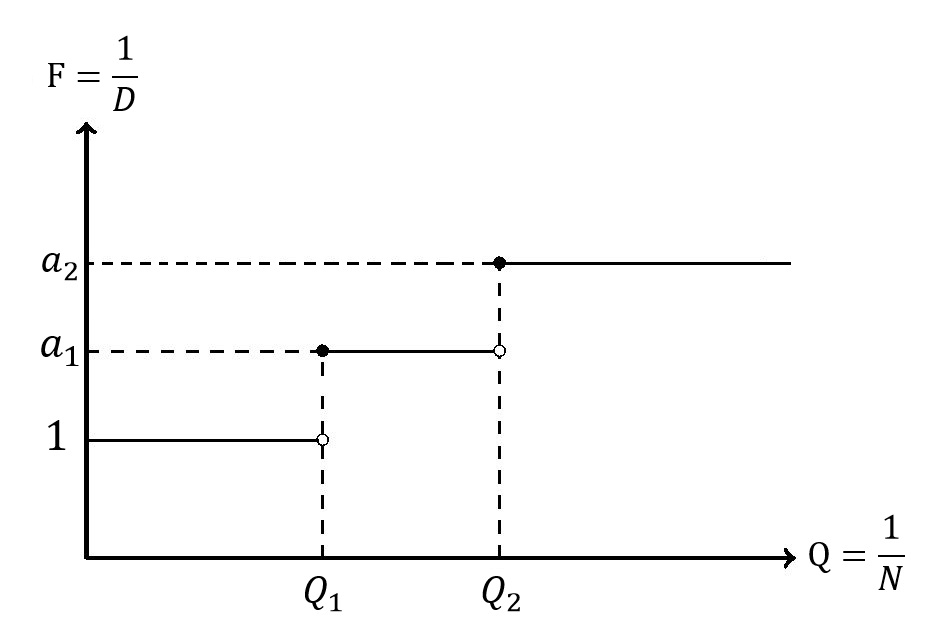}
    \caption{The Staircase fidelity-quality profile with $K=2$.}
\end{figure}
\subsubsection{$Q_k=\gamma^k$, $a_k=\lambda^k$ and $K=\infty$}
In this part, we assume that $Q_k=\gamma^k$, $a_k=\lambda^k$ where $\gamma$ and $\lambda$ are positive constants and $K=\infty$ .We find the lower and upper bounds for the profile as follows.
\paragraph{lower Bound for $E_{min}(\cal{D})$}
We begin with the lower bounding $E_{min}(\cal{D})$ by the following Theorem.
\begin{theorem}
For a staircase profile with $Q_k=\gamma^k$, $a_k=\lambda^k$ and $K=\infty$, the minimum required energy is lower-bounded as 
\[
E_{\min}({\cal D})\geq \frac{\log \frac{\lambda}{4}}{\gamma-1}. 
\]
\end{theorem}
\begin{IEEEproof}
Invoking Lemma~\ref{Lemma1} by choosing $\tau_k={\cal D}(N_k)$ for $k=1,\ldots,K-1$ in (\ref{eqtn:lemma1}), where $N_1\geq \ldots\geq N_K$ to be chosen later and ${\cal D}(N) = \frac{1}{{\cal F}\left(\frac{1}{N}\right)}$
as before, we obtain
\begin{eqnarray*}
\lefteqn{E_{\min}({\cal D})}\\
& \geq & N_1\log\frac{1+{\cal D}(N_1)}{2{\cal D}(N_1)}  \\
&& +\sum_{k=2}^{K}N_{k}\log\frac{\left[1+{\cal D}(N_k)\right]\left[{\cal D}(N_k)+{\cal D}(N_{k-1})\right]}{2[1+{\cal D}(N_{k-1})]{\cal D}(N_k)}\\
 & \geq & N_1\log\frac{1}{2{\cal D}(N_1)}+\sum_{k=2}^{K}N_k\log\frac{{\cal D}(N_{k-1})}{4{\cal D}(N_k)}\\
 & \geq & \sum_{k=1}^{K}N_{k}\log\left(\frac{{\cal D}(N_{k-1})}{4{\cal D}(N_k)}\right).
\end{eqnarray*}
Now, if we substitue $N_k=\gamma^{-k}$ and $D(N_k)=\frac{1}{a_k}=\lambda^{-k}$
for $k=1,\ldots,K$, it follows that
\begin{align}
    E_{\min}({\cal D}) & \geq \sum_{k=1}^K \gamma^{-k} \log \frac{\lambda}{4} \nonumber\\ 
& = \log \frac{\lambda}{4}\sum_{k=1}^K \gamma^{-k}
\end{align}
for any $K\geq 1$. The lower bound will then be obtained in (\ref{lowerb}) as $K\rightarrow\infty$.
\begin{align}
\label{lowerb}
    E_{\min}({\cal D}) & \geq \frac{\log \frac{\lambda}{4}}{\gamma-1}.
\end{align}
\end{IEEEproof}

\paragraph{Upper Bound for $E_{min}(\cal{D})$}
In this part, we apply coding scheme of \cite{c9} which is for square-law profile to our staircase profile. In Figure 2, the fidelity-quality tradeoff of our scheme and staircase profile is shown.
\begin{figure}
  \centering
    \includegraphics[width=0.5\textwidth]{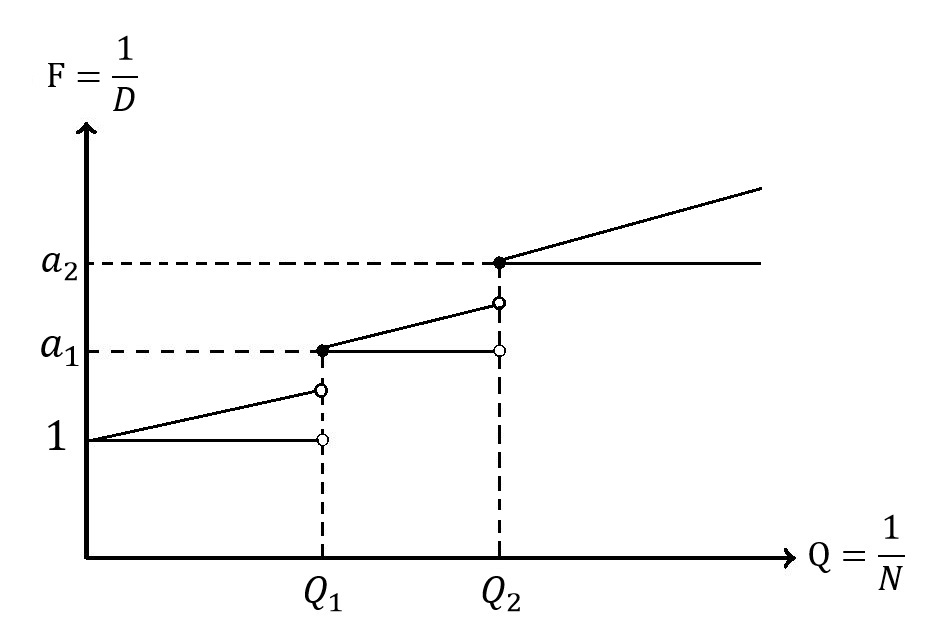}
    \caption{The achieved fidelity F(Q) versus Staircase profile with $K=2$.}
\end{figure}
We find upper bounds for the minimum energy based on two different coding schemes.

In \cite{c9}, the authors used to send uncoded energy $E_0$ at first layer and then just sending digital energy for the other layers. 
Since following the staircase profile is similar to following square-law profile, the first coding scheme which comes to mind is using just digital coding which is the especial case of \cite{c9} proposed square-law profile scheme with $E_0=0$.
In Figure 2, each piece-wise line has slope equal to $E_0$. If we set $E_0=0$, then $F$ and staircase profile are totally matched which means the profile is achieved.
Thus, the result is illustrated as following Theorem.
\begin{theorem}
For a staircase profile with $Q_k=\gamma^k$, $a_k=\lambda^k$ and $K=\infty$, the minimum required energy is upper-bounded as 
\begin{align}
    E_{\min}({\cal D}) & \leq \frac{\log \lambda}{\gamma-1}.\nonumber
\end{align}
\end{theorem}
\begin{IEEEproof}
By putting $E_0=0$ in equations discussed in \cite{c9},
the minimum energy for this scheme could be
\begin{align}\label{emin}
E_{min}(\cal{D})&\leq \sum_{k=1}^{K} B_k \nonumber\\
E_{min}(\cal{D})& \leq \sum_{k=1}^{K} N_k\log (\frac{a_k}{a_{k-1}}),
\end{align}
where $B_k$ is digital transmission energy corresponding to layer $k$.
Now, substituting $N_k=\gamma^{-k}$ and $a_k=\lambda^{k}$
for $k=1,\ldots,K$, it follows that
\begin{align}
    E_{\min}({\cal D}) & \leq \sum_{k=1}^K \gamma^{-k} \log \lambda \nonumber\\ 
& =  \log \lambda\sum_{k=1}^K \gamma^{-k}
\end{align}
for any $K\geq 1$. The lower bound will then be obtained in (\ref{upperb}) as $K\rightarrow\infty$.
\begin{align}
\label{upperb}
    E_{\min}({\cal D}) & \leq \frac{\log \lambda}{\gamma-1}.
\end{align}
\end{IEEEproof}
Now we consider our second scheme in which we can send uncoded or analog energy in the first layer and then only using the digital energy for other layers. It is somehow similar to square-law profile analysis by having $E_0$ nonzero. 
Therefore,
\begin{align}
B_k=N_k\log\frac{E_0 Q_k+ \beta_k}{E_0 Q_k+ \beta_{k-1}},\nonumber
\end{align}
where $\beta_{k}=\frac{1}{\sigma^2_{S_k}}$ and $S_k$ is the quantization error from layer $k$, respectivly.
Meanwhile, according to Figure 2, we should have this constraint in order to follow the staircase profile.
\begin{align}
E_0 Q_k +\beta_k \ge a_k,\ k= 1,2,..., K.\nonumber
\end{align}
Our goal is to minimize the total energy including uncoded and digital energy. First, we assume $E_0$ is fixed and minimize the total digital energy denoted by $E_D$. To do this, we deal with the optimization problem over $\beta_k$ as follows
\begin{align}\label{optprob}
&\min E_D=\sum_{k=1}^{K} B_k\nonumber\\
&s.t.: \beta_k \ge a_k-E_0 Q_k \ k=1,2,...,K.
\end{align}
\begin{lemma}
$E_D(\beta_1,\beta_2,...,\beta_K)$ is an increasing and concave function with respect to $(\beta_1,\beta_2,...,\beta_K)$.
\end{lemma}
\begin{IEEEproof}
First, we find the gradient vector of $E_D(\beta_1,\beta_2,...,\beta_K)$ as follows
\begin{align}
\bigtriangledown E_D(\beta_1,\beta_2,...,\beta_K)&=
  \begin{bmatrix}
    \frac{N_1}{E_0 Q_1 + \beta_1}-\frac{N_2}{E_0 Q_2 + \beta_1}\\
    . \\
    . \\
    . \\
    \frac{N_k}{E_0 Q_k + \beta_k}-\frac{N_{k+1}}{E_0 Q_{k+1} + \beta_k}\\
    . \\
    . \\
    . \\
    \frac{N_K}{E_0 Q_K + \beta_K}
    \end{bmatrix}.\nonumber
    \end{align}
Each element of gradient vector except the last one (for $i=1,2,..,K-1$) is equal to 
\begin{align}
\frac{\partial  E_D}{\partial \beta_k}&=
\frac{N_k}{E_0 Q_k +\beta_k}-\frac{N_{k+1}}{E_0 Q_{k+1} +\beta_k}\nonumber\\
&=\frac{E_0 (N_k Q_{k+1}-N_{k+1}Q_k)+\beta_k (N_k-N_{k+1})}{(E_0 Q_k+\beta_k)(E_0 Q_{k+1}+\beta_k)}.
\end{align}
Since $Q_{k+1}> Q_k$ and $N_{k+1} <  N_k $, we conclude that $N_k Q_{k+1}-N_{k+1}Q_k >0$. It is obvious that the last element of this vector is positive.
Therefore, each element in gradient vector is positive and thus the function is increasing.

Now, we calculate the Hessian matrix of $E_D(\beta_1,\beta_2,...,\beta_K)$ in (\ref{hessianmatrix}).
%
    \begin{figure*}
    \begin{equation}\label{hessianmatrix}
    \bigtriangledown^2 E_D(\beta_1,\beta_2,...,\beta_K)
=
  \begin{bmatrix}
    -\frac{N_1}{(E_0 Q_1 + \beta_1)^2}+\frac{N_2}{(E_0 Q_2 + \beta_1)^2}& 0 & . & . & . & 0\\
    0& . & . & . & . & . \\
    . & . &. &. &. & .\\
    . \\
    0 & . &. & -\frac{N_k}{(E_0 Q_k + \beta_{k})^2}+\frac{N_{k+1}}{(E_0 Q_{k+1} + \beta_k)^2}
     & . & 0 \\
    . \\
    . & . &. &. &. & 0\\
    0 & . &. & .& 0 & -\frac{N_K}{(E_0 Q_K + \beta_K)^2}
    \end{bmatrix}
    \end{equation}
\end{figure*}
Each element of Hessian matrix except the last one (for $i=1,2,..,K-1$) is equal to 
\begin{align}
\frac{\partial  E^2_D}{\partial \beta^2_k}&=
\frac{N_{k+1}}{(A_0 Q_{k+1} +\beta_k)^2}-\frac{N_{k}}{(A_0 Q_{k} +\beta_k)^2}\nonumber\\
&=\frac{N_{k+1}(A_0 Q_{k}+\beta_k)^2-N_k(A_0 Q_{k+1}+\beta_k)^2}{(A_0 Q_k+\beta_k)^2(A_0 Q_{k+1}+\beta_k)^2}.
\end{align}
Since $N_{k+1} <  N_k$, $Q_{k+1}> Q_k$ and $N_{k+1}(A_0 Q_{k}+\beta_k)^2-N_k(A_0 Q_{k+1}+\beta_k)^2<0$. It is clear that the last element of this matrix is also negative.
Thus, each diagonal element in Hessian matrix is negative and $E_D(\beta_1,\beta_2,...,\beta_K)$ is a concave function.
\end{IEEEproof}
According to that $E_D(\beta_1,\beta_2,...,\beta_K)$ is a concave function and since we want to minimize it, the optimal solution is where the constraint of optimization problem (\ref{optprob}) is satisfied with equality which sates $\beta_k=a_k-E_0 Q_k$. Therefore, the total energy for digital transmission is as follows
\begin{align}
E_D^*&=E_D(\beta_k=a_k-E_0 Q_k) \nonumber\\
&=\sum_{k=1}^{K} N_k\log\frac{a_k}{E_0 Q_k+a_{k-1}-E_0 Q_{k-1}}\nonumber\\
&=\sum_{k=1}^{K} N_k\log\frac{a_k}{E_0(\frac{1}{N_k}-\frac{1}{N_{k-1}}) +a_{k-1}}.
\end{align}
So far, we minimized the total energy for digital transmission. Now, we plan to find the minimum of total energy which also includes analog part of transmission. Thus, we add $E_0$ to $E^*_D$.
\begin{align}
\label{etot}
E_{min}(\cal{D})&=E_0+E_D^*\nonumber\\
&=E_0+\sum_{k=1}^{K} N_k\log\frac{a_k}{E_0(\frac{1}{N_k}-\frac{1}{N_{k-1}}) +a_{k-1}}.
\end{align}
\begin{rmrk}
Please note that if $E_0=0$, the total minimum energy could be 
 $\sum_{k=1}^{K} N_k\log (\frac{a_k}{a_{k-1}})$
which exactly matches with (\ref{emin}) presented before in this section.
\end{rmrk}
Our goal is to minimize the $E_{min}(\cal{D})$ over $E_0$ with respect to the following constraint
\begin{align}
\beta_k > \beta_{k-1} > ...> \beta_1 >1.
\end{align}
This constraint is equivalent to the following constraints for $E_0$,
\begin{align}
\beta_1 > 1 &\equiv a_1-E_0 Q_1 >1 	\Rightarrow E_0 < N_1 (a_1 -1),\nonumber\\
\beta_k > \beta_{k-1} &\equiv a_k-\frac{E_0}{N_k} > a_{k-1}-\frac{E_0}{N_{k-1}},\nonumber\\
E_0 &< \frac{(a_k-a_{k-1})}{(\frac{1}{N_k}-\frac{1}{N_{k-1}})}.
\end{align} 
By combining the above constraints, we conclude that
\begin{align}
\label{C}
E_0 & < \min\bigg\{N_1 (a_1 -1),\frac{\big(a_k-a_{k-1}\big)}{\big(\frac{1}{N_k}-\frac{1}{N_{k-1}}\big)}\bigg\},\nonumber\\
k & =1,2,...,K.
\end{align} 
Therefore, the optimization problem to minimize the total energy is
\begin{align}
\label{optstair}
&\min  (\ref{etot}) \nonumber\\
&s.t.: (\ref{C}).
\end{align}
\begin{lemma}
The objective function in (\ref{optstair}) which is defined in (\ref{etot}) is convex with respect to $E_0$.
\end{lemma}
\begin{IEEEproof}
We calculate the second derivative of $E_{min}(\cal{D})$ with respect to $E_0$.
\begin{align}
E_{min}(\cal{D})&=E_0+\sum_{k=1}^{K} N_k\log\frac{a_k}{E_0(\frac{1}{N_k}-\frac{1}{N_{k-1}}) +a_{k-1}}\nonumber\\
\frac{\partial E_{min}(\cal{D})}{\partial E_0}&=1-\sum_{k=1}^{K} N_k \frac{(\frac{1}{N_k}-\frac{1}{N_{k-1}})}{E_0 (\frac{1}{N_k}-\frac{1}{N_{k-1}})+a_{k-1}}\nonumber\\
\frac{\partial E^2_{min}(\cal{D})}{\partial E_0}&=\sum_{k=1}^{K} N_k(\frac{1}{N_k}-\frac{1}{N_{k-1}})^2 \frac{1}{(E_0 (\frac{1}{N_k}-\frac{1}{N_{k-1}})+a_{k-1})^2}\nonumber\\
& \ge 0.
\end{align}
Since the second derivative of $E_{min}(\cal{D})$ with respect to $E_0$ is always non-negative, $E_{min}(\cal{D})$ is convex with respect to $E_0$.
\end{IEEEproof}
Now, finding the optimal solution for problem (\ref{optstair}) is straight forward. First, we set $\frac{\partial E_{min}(\cal{D})}{\partial E_0}=0$. We denote the $E_0$ which is the solution of this equation as $E^*_0$.

Therefore, We have
\begin{equation}
\label{generalstar}
    E^*_{min}(\cal{D})=
    \begin{cases}
       E_{min}(E_0=Z), & \text{if}\ Z \leq E^*_0 \\
      E_{min}(E^*_0), & \text{if}\ Z > E^*_0 
    \end{cases}
 \end{equation}
 where $Z=\min\bigg\{N_1 (a_1 -1),\frac{(a_k-a_{k-1})}{(\frac{1}{N_k}-\frac{1}{N_{k-1}})}\bigg\}$ over $k=1,...,K$.
Please note that our analysis until this point is true for any $K$, $a_k$ and $Q_k$.
Now, if we substitute $Q_k=\gamma^k$, $a_k=\lambda^k$ and $K=\infty$ we get
\begin{align}
    Z&=\min\bigg\{\gamma^{-1} (\lambda-1),\frac{\lambda^k-\lambda^{k-1}}{\gamma^k-\gamma^{k-1}}\bigg\},\nonumber\\
    Z&=(\lambda-1)\min\bigg\{\gamma^{-1} ,\frac{(\frac{\lambda}{\gamma})^{k-1}}{\gamma-1}\bigg\},\nonumber
    \end{align}
    for $k=1,2,...,\infty$. Thus,
    \begin{align}
    Z=
    \begin{cases}
       \frac{\lambda-1}{\gamma}, & \text{if}\ \frac{\lambda}{\gamma} \geq 1 \\
      0, & \text{if}\ \frac{\lambda}{\gamma} < 1 
    \end{cases}
\end{align}

\subsubsection{$K=2$}
In this part, we assume that $K=2$ and find the general lower and upper bounds for the profile.

\paragraph{Lower Bound for $E_{min}(\cal D)$}
We begin with the lower bounding $E_{min}(\cal{D})$ by the following Theorem.
\begin{theorem}
For a staircase profile with $K=2$, the minimum required energy is lower-bounded as 
\begin{align}
E_{\min}({\cal D})\geq E^*_{l,min}. 
\end{align}
where $E^*_{l,min}$ is defined in (\ref{Eminstar}).
\end{theorem}
\begin{figure*}
 \begin{equation}
\label{Eminstar}
   E^*_{l,min}=
    \begin{cases}
       N_1 \log \frac{a_1(a_1a_2-a_2-a_1+1)(N_1-N_2)}{N_1(a_1a_2-a_2+a_1-a_1^2)} + N_2 \log \frac{N_2(-a_1a_2+a_1-a_2+a_2^2)}{(a_1a_2-a_2-a_1+1)(N_1-N_2)} & \text{if}\ \frac{a_1-1}{a_2-1}<\frac{N_2}{N_1}<\frac{a_2(a_1-1)}{a_1(a_2-1)} \\
      N_2 \log a_2 & \text{if}\ \frac{N_2}{N_1}>\frac{a_2(a_1-1)}{a_1(a_2-1)} \\
      N_1 \log a_1 & \text{if}\ \frac{N_2}{N_1}<\frac{a_1-1}{a_2-1}
    \end{cases}
 \end{equation}
 \end{figure*}

\begin{IEEEproof}
Invoking Lemma~\ref{Lemma1} by choosing $K=2$, $D(N_k)=\frac{1}{a_k}$, and $\tau_2=0$
 the lower bound is achieved as
 \begin{align}
\label{tow1}
E_{min}(\cal{D})&\ge N_1 \log \frac{1+\tau_1}{\frac{1}{a_1}+\tau_1}\nonumber\\
&+ N_2 \log \frac{(\frac{1}{a_2}+\tau_{1})}{(1+\tau_{1})(\frac{1}{a_2})}.
\end{align}
By maximizing the right hand side of (\ref{tow1}) with respect to $\tau_1 \ge 0$, we get

\begin{equation}
\label{towstar}
   \tau^*_1=
    \begin{cases}
       \frac{N_1(\frac{1}{a_1}-1)\frac{1}{a_{2}}-N_{2}(\frac{1}{a_{2}}-1)\frac{1}{a_1}}{N_1(1-\frac{1}{a_1})-N_{2}(1-\frac{1}{a_{2}})} & \text{if}\ \frac{a_1-1}{a_2-1}<\frac{N_2}{N_1}<\frac{a_2(a_1-1)}{a_1(a_2-1)} \\
      \infty & \text{if}\ \frac{N_2}{N_1}>\frac{a_2(a_1-1)}{a_1(a_2-1)} \\
      0 & \text{if}\ \frac{N_2}{N_1}<\frac{a_1-1}{a_2-1}
    \end{cases}
 \end{equation}
 By substituting (\ref{towstar}) in (\ref{tow1}), we reach to the lower bound in (\ref{Eminstar}).
 \end{IEEEproof}
\paragraph{Upper Bound for $E_{min}(\cal D)$}
We continue our analysis with the upper bounding $E_{min}(\cal D)$ by the following Theorem.
\begin{theorem}
For a staircase profile with $K=2$, the minimum required energy is upper-bounded as 
\begin{align}
E_{\min}({\cal D})\leq E^*_{u,min}. 
\end{align}
where $E^*_{u,min}$ is defined in (\ref{upstar}).
\end{theorem}
\begin{IEEEproof}
When we discussed our method to find the upper bound in previous sub-section, we found the general solution as in (\ref{generalstar}) which is true for any $K$.
For especial case, when $K=2$, we can find the closed form solution for $E^*_0$ as following.
{\small
\begin{align}
 E_{min}(\cal{D})&=E_0 +N_1 \log \frac{1}{\frac{E_0}{a_1 N_1}+\frac{1}{a_1}}+N_2 \log \frac{1}{\frac{E_0}{a_2}(\frac{1}{N_2}-\frac{1}{N_1})+\frac{a_1}{a_2}}\nonumber\\
\frac{\partial E_{min}(\cal{D})}{\partial E_0}&=1-\frac{N_1}{E_0 + N_1}-\frac{N_2}{E_0+\frac{\frac{a_1}{a_2}}{\frac{1}{a_2}(\frac{1}{N_2}-\frac{1}{N_1})}}=0\nonumber\\
	&\Rightarrow E_0^2+E_0(M-N_2)-N_2 N_1=0\nonumber\\
    &\Rightarrow E^*_0=\frac{-(M-N_2)+\sqrt[]{(M-N_2)^2+4N_2N_1}}{2}
\end{align}}
where $M=\frac{a_1}{(\frac{1}{N_2}-\frac{1}{N_1})}$. Thus,
\begin{equation}
\label{upstar}
    E^*_{u,min}=
    \begin{cases}
       E_{min}(E_0=L), & \text{if}\ L \leq E^*_0 \\
      E_{min}(E^*_0), & \text{if}\ L > E^*_0 
    \end{cases}
 \end{equation}
 where $L=\min\bigg\{N_1 (a_1-1),\frac{(a_2-a_{1})}{(\frac{1}{N_2}-\frac{1}{N_{1}})}\bigg\}$.
 \end{IEEEproof}
\section{Conclusions and Future Work}

Minimum energy required to achieve a distortion-noise profile, was studied for robust transmission of Gaussian sources over Gaussian channels.
In order to analyze the minimum energy behavior for the staircase distortion noise profile, the lower and upper bounds were proposed by our coding schemes. 

For future, we are interested to study distortion-noise profile problem in Multiple Access Channels (MAC). In MAC, instead of having one distortion function, we deal with at least two distortion functions and distortion regions. 


\end{document}